\documentclass[twocolumn,superscriptaddress,showpacs,floatfix,prl]{revtex4}
\usepackage{amsmath}
\usepackage{graphicx}

\begin{document}

\title{Determination of the stiffness of the nuclear symmetry energy from
isospin diffusion}
\author{Lie-Wen Chen}
\thanks{Current Address: Institute of Theoretical Physics, Shanghai Jiao
Tong University, Shanghai 200240, China}
\affiliation{Cyclotron Institute and Physics Department, Texas A\&M University, College
Station, Texas 77843-3366}
\author{Che Ming Ko}
\affiliation{Cyclotron Institute and Physics Department, Texas A\&M University, College
Station, Texas 77843-3366}
\author{Bao-An Li}
\affiliation{Department of Chemistry and Physics, P.O. Box 419, Arkansas State
University, State University, Arkansas 72467-0419}
\date{\today }

\begin{abstract}
With an isospin- and momentum-dependent transport model, we find that the
degree of isospin diffusion in heavy ion collisions at intermediate energies
is affected by both the stiffness of the nuclear symmetry energy and the
momentum dependence of the nucleon potential. Using a momentum dependence
derived from the Gogny effective interaction, recent experimental data from
NSCL/MSU on isospin diffusion are shown to be consistent with a nuclear
symmetry energy given by $E_{\text{sym}}(\rho )\approx 31.6(\rho
/\rho_{0})^{1.05} $ at subnormal densities. This leads to a significantly
constrained value of about $-550$ MeV for the isospin-dependent part of the
isobaric incompressibility of isospin asymmetric nuclear matter.
\end{abstract}

\maketitle

Knowledge on the density dependence of nuclear symmetry energy is important
for understanding not only the structure of radioactive nuclei \cite%
{oya,brown,hor01,furn02} but also many important issues in nuclear
astrophysics \cite{bethe,lat01}, such as nucleosynthesis during
pre-supernova evolution of massive stars and the cooling of protoneutron
stars. Although the nuclear symmetry energy at normal nuclear matter density
is known to be around $30$ \textrm{MeV} from the empirical liquid-drop mass
formula \cite{myers,pomorski}, its values at other densities are poorly
known \cite{ireview98,ibook}. Studies based on either microscopic many-body
theories or phenomenological approaches have so far given widely divergent
predictions on the density dependence of nuclear symmetry energy \cite{bom}.
Empirically, the incompressibility of asymmetric nuclear matter is
essentially undetermined \cite{shlomo93}, even though the incompressibility
of symmetric nuclear matter at its saturation density $\rho _{0}=0.16$ fm$%
^{-3}$ has been determined to be $231\pm 5$ MeV from nuclear giant monopole
resonances \cite{youngblood99} and the equation of state at densities of $%
2\rho_0<\rho<5\rho_0$ has been constrained by measurements of collective
flows in nucleus-nucleus collisions \cite{pawel02}.

In light of the new opportunities provided by radioactive beams, a lot of
interests and activities have recently been devoted to extract information
on the density dependence of nuclear symmetry energy from reactions induced
by such nuclei \cite%
{li97,fra1,fra2,xu00,tan01,bar02,betty,lis,li00,npa01,li02,chen}. In
particular, isospin diffusion in heavy-ion collisions is found to depend
sensitively on the density dependence of nuclear symmetry energy \cite%
{farine91,shi03,li04prc}. Within a momentum-independent transport model, in
which the nuclear potential depends only on local nuclear density, the
isospin diffusion data from recent experiments at the NSCL/MSU (National
Superconducting Cyclotron Laboratory at Michigan State University) was found
to favor a quadratic density dependence for the interaction part of nuclear
symmetry energy \cite{tsang04}. This conclusion has stimulated much interest
because of its implications to nuclear many-body theories and nuclear
astrophysics. However, the nuclear potential acting on a nucleon is known to
depend also on its momentum. For nuclear isoscalar potential, its momentum
dependence is well-known and is important in extracting information about
the equation of state of symmetric nuclear matter \cite%
{gbd87,welke88,gale90,pan93,zhang94,greco99,dan00,persram02,pawel02}. Very
recently, the momentum-dependence of the isovector (symmetry) potential \cite%
{bom,hod94,das03,li04mass} was also shown to be important for understanding
a number of isospin related phenomena in heavy-ion reactions \cite%
{li04a,rizzo04,chen04}. It is thus necessary to include momentum dependence
in both the isoscalar and isovector potentials for studying the effect of
nuclear symmetry energy on isospin diffusion. In this Letter, we shall show
that the isospin diffusion data are consistent instead with a softer
symmetry energy that is nearly linear in density within the
momentum-dependent effective interaction used in present study.

Our study is based on an isospin-dependent transport model IBUU04 that uses
experimental nucleon-nucleon cross sections in free space and includes the
momentum dependence in both the isoscalar and isovector potentials \cite%
{li04a}. Although the momentum dependence of the isoscalar potential is
known empirically, that of the isovector potential is not as well determined 
\cite{baran04}. In the IBUU04 model, it is based on the Gogny effective
interactions. Specifically, the potential $U(\rho ,\delta ,\mathbf{p},\tau )$
for a nucleon with isospin $\tau $ ($1/2$ for neutrons and $-1/2$ for
protons) and momentum $\mathbf{p}$ in asymmetric nuclear matter at total
density $\rho $ is given by \cite{das03} 
\begin{eqnarray}
U_{\text{MDI}}(\rho ,\delta ,\mathbf{p},\tau ) &=&A_{u}\frac{\rho _{\tau
^{\prime }}}{\rho _{0}}+A_{l}\frac{\rho _{\tau }}{\rho _{0}}+B\left( \frac{%
\rho }{\rho _{0}}\right) ^{\sigma }(1-x\delta ^{2})  \notag \\
&-&8\tau x\frac{B}{\sigma +1}\frac{\rho ^{\sigma -1}}{\rho _{0}^{\sigma }}%
\delta \rho _{\tau ^{\prime }}  \notag \\
&+&\frac{2C_{\tau ,\tau }}{\rho _{0}}\int d^{3}\mathbf{p}^{\prime }\frac{%
f_{\tau }(\mathbf{r},\mathbf{p}^{\prime })}{1+(\mathbf{p}-\mathbf{p}^{\prime
})^{2}/\Lambda ^{2}}  \notag \\
&+&\frac{2C_{\tau ,\tau ^{\prime }}}{\rho _{0}}\int d^{3}\mathbf{p}^{\prime }%
\frac{f_{\tau ^{\prime }}(\mathbf{r},\mathbf{p}^{\prime })}{1+(\mathbf{p}-%
\mathbf{p}^{\prime })^{2}/\Lambda ^{2}},  \label{mdi}
\end{eqnarray}%
where $\rho _{\tau }$ and $\rho _{\tau }^{\prime }$ denote proton or neutron
density with $\tau \neq \tau ^{\prime }$; and $\delta \equiv (\rho _{n}-\rho
_{p})/\rho $ is the isospin asymmetry. The $f_{\tau }(\mathbf{r},\mathbf{p})$
denotes the phase-space distribution function at coordinate $\mathbf{r}$ and
momentum $\mathbf{p}$. The corresponding momentum-dependent interaction
(MDI) leads to an incompressibility of $K_{0}=211$ MeV for the symmetric
nuclear matter at saturation density. On the right hand side of Eq.(\ref{mdi}%
), the first four terms with $\sigma =3/4$ and $B=106.35$ MeV describe the
momentum-independent interaction. The terms with parameters $C_{\tau ,\tau
}=-11.7$ MeV and $C_{\tau ,\tau ^{\prime }}=-103.4$ MeV describe the
momentum-dependent interaction of a nucleon of isospin $\tau $ and momentum $%
\mathbf{p}$ with like and unlike nucleons in the background fields,
respectively. With the parameter $\Lambda =1.0p_{F}^{0}$, where $p_{F}^{0}$
denotes nucleon Fermi momentum at $\rho _{0}$, the isoscalar potential $%
(U_{n}(\rho ,p)+U_{p}(\rho ,p))/2$ coincides with predictions from the
variational many-body theory using inputs constrained by nucleon-nucleon
scattering data \cite{li04a}, and the isovector potential $(U_{n}(\rho
,p)-U_{p}(\rho ,p))/2$ also agrees with the momentum dependence of the Lane
potential extracted from low-energy nucleon-nucleus scattering experiments 
\cite{li04a} and recent neutron-nucleus data at 96 MeV \cite{klug03}.

\begin{figure}[th]
\includegraphics[scale=0.65]{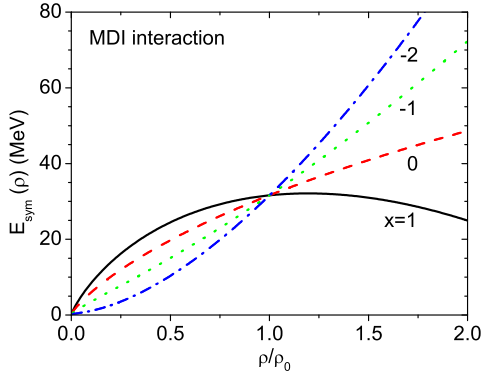}
\caption{{\protect\small (Color online) Density dependence of nuclear
symmetry energy for different values of $x$ parameter in Eq.(\protect\ref%
{mdi}).}}
\label{Esym}
\end{figure}

The parameter $x$ in Eq.(\ref{mdi}) is introduced to mimic the various
theoretical predictions for the density dependence of the nuclear symmetry
energy $E_{\text{sym}}(\rho )$, which is defined via the parabolic
approximation to the nucleon specific energy in an isospin asymmetric
nuclear matter \cite{ireview98,ibook}, i.e., 
\begin{equation}
E(\rho ,\delta )=E(\rho ,\delta =0)+E_{\text{sym}}(\rho )\delta ^{2}+%
\mathcal{O}(\delta ^{4}).  \label{eos}
\end{equation}%
With $x=1$, for instance, the $E_{\text{sym}}(\rho )$ is what predicted by a
Hartree-Fock calculation using the Gogny effective interaction \cite{das03}.
The parameters $A_{l}(x)$ and $A_{u}(x)$ are $A_{l}(x)=-120.57+2Bx/(\sigma
+1)$ and $A_{u}(x)=-95.98-2Bx/(\sigma +1)$, respectively. Including also the
well-known contribution from nucleon kinetic energies, i.e., $E_{\text{sym}%
}^{\text{kin}}(\rho )=(2^{2/3}-1)\frac{3}{5}E_{F}^{0}(\rho /\rho
_{0})^{2/3}\approx 13.0(\rho /\rho _{0})^{2/3}$, the density dependence of
nuclear symmetry energy is shown in Fig. \ref{Esym} for $x=1$, $0$, $-1$ and 
$-2$.

\begin{table}[tbp]
\caption{{\protect\small The parameters }$F${\protect\small \ (MeV), }$G$ 
{\protect\small , }$K_{\text{sym}}${\protect\small \ (MeV), } $L$%
{\protect\small \ (MeV), and }$K_{\text{asy}}${\protect\small \ (MeV) for
different values of} $x${\protect\small .}}
\label{MDIx}%
\begin{tabular}{ccccccc}
\hline\hline
$x$ & \quad $F$ & $G$ & $K_{\text{sym}}$ & $L$ & $K_{\text{asy}}$ &  \\ 
\hline
$1$ & $107.232$ & $1.246$ & $-270.4$ & $16.4$ & -368.8 &  \\ 
$0$ & $129.981$ & $1.059$ & $-88.6$ & $62.1$ & -461.2 &  \\ 
$-1$ & $3.673$ & $1.569$ & $94.1$ & $107.4$ & -550.3 &  \\ 
$-2$ & $-38.395$ & $1.416$ & $276.3$ & $153.0$ & -641.7 &  \\ \hline\hline
\end{tabular}%
\end{table}

The interaction part of nuclear symmetry energy can be parameterized by $E_{%
\text{sym}}^{\mathrm{pot}}(\rho ) =F(x)\rho /\rho_{0} +(18.6-F(x))(\rho
/\rho _{0})^{G(x)}$, with $F(x)$ and $G(x)$ given in Table \ref{MDIx} for $%
x=1$, $0$, $-1$ and $-2$. Also shown are other characteristics of the
symmetry energy, including its slope $L\equiv 3\rho_0\left(dE_{\text{sym}%
}/d\rho\right)_{\rho=\rho_0}$ and curvature $K_{\text{sym}}\equiv
9\rho_0^{2} \left(d^{2}E_{\text{sym}}/d\rho ^{2}\right)_{\rho=\rho_0}$ at $%
\rho_0$, as well as the isospin-dependent part $K_{\mathrm{asy}}\approx K_{%
\mathrm{sym}}-6L$ \cite{bar02} of the isobaric incompressibility of
asymmetric nuclear matter $K(\delta)=K_{0}+K_{\text{asy}}\delta^{2}$ \cite%
{prakash85,lopez88}.

For comparisons we have also constructed the following momentum-independent
potential (SBKD) that has $K_{0}=200$ MeV \cite{sbkd} and exactly the same $%
E_{\text{sym}}(\rho )$ as the MDI interaction: 
\begin{eqnarray}
U_{\text{SBKD}}(\rho ,\delta ,\tau )&\equiv&-356~\rho /\rho _{0}+303~(\rho
/\rho _{0})^{7/6}  \notag \\
&+&4\tau E_{\text{sym}}^{\mathrm{pot}}(\rho )+(18.6-F(x))  \notag \\
&\times&(G(x)-1)(\rho /\rho _{0})^{G(x)}\delta ^{2}.  \label{Usbkd}
\end{eqnarray}

Isospin diffusion in heavy ion collisions can in principle be studied by
examining the average isospin asymmetry of the projectile-like residue in
the final state. Since reactions at intermediate energies are complicated by
preequilibrium particle emission and production of neutron-rich fragments at
mid-rapidity, differences of isospin diffusions in mixed and symmetric
systems are usually used to minimize these effects \cite{tsang04}. To study
isospin diffusion in $^{124}$Sn + $^{112}$Sn reactions at $E=50$ \textrm{%
MeV/nucleon} and an impact parameter of $b=6$ fm, we thus also consider the
reaction systems $^{124}$Sn + $^{124}$S and $^{112}$Sn + $^{112}$Sn and $%
^{124}$Sn + $^{112}$Sn at same energy and impact parameter as in Ref.\cite%
{tsang04}. The degree of isospin diffusion in the reaction $^{124}$Sn + $%
^{112}$Sn is then measured by \cite{rami00} 
\begin{equation}
R_{i}=\frac{2X_{^{124}\text{Sn}+^{112}\text{Sn}}-X_{^{124}\text{Sn}+^{124}%
\text{Sn}}-X_{^{112}\text{Sn}+^{112}\text{Sn}}}{X_{^{124}\text{Sn}+^{124}%
\text{Sn}}-X_{^{112}\text{Sn}+^{112}\text{Sn}}}  \label{Ri}
\end{equation}%
where $X$ is the average isospin asymmetry $\left\langle \delta
\right\rangle $ of the $^{124}$Sn-like residue defined as the composition of
nucleons with local densities higher than $\rho _{0}/20$ and velocities
larger than $1/2$ the beam velocity in the c.m. frame. A density cut of $%
\rho _{0}/8$ is found to give almost same results. In ideal case, the value
of $R_{i}$ ranges between $0.05$ and $1$ from complete mixing to full
transparency.

\begin{figure}[th]
\includegraphics[scale=0.75]{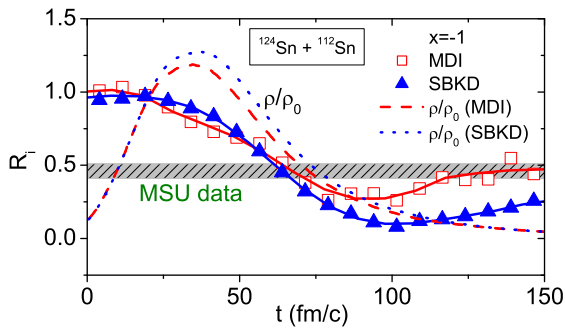}
\caption{{\protect\small (Color online) Time evolutions of }$R_{i}$%
{\protect\small \ and average central density for MDI and SBKD interactions
with }$x=-1$.}
\label{RiTime}
\end{figure}

\begin{figure}[th]
\includegraphics[scale=0.75]{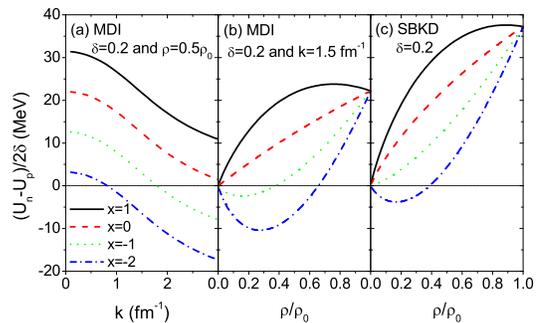}
\caption{{\protect\small (Color online) Symmetry potential as a function of
momentum (a) or density (b) with the MDI interaction and SBKD interaction
(c).}}
\label{Upot}
\end{figure}

As an example of present more realistic calculations, we show in Fig.\ \ref%
{RiTime} the time evolutions of $R_{i}$ and average central density
calculated with $x=-1$ using both MDI and SBKD interactions. It is seen that
the isospin diffusion process occurs mainly from about $30$ fm/c to $80$
fm/c corresponding to average central density from about $1.2\rho _{0}$ to $%
0.3\rho _{0}$. However, the value of $R_{i}$ still changes slightly with
time until after about $120$ fm/c when projectile-like and target-like
residues are well separated. This is partly due to the fact that the
isovector potential remains appreciable at low density as shown in Fig.\ \ref%
{Upot}, where the symmetry potential $(U_{n}-U_{p})/2\delta $ is shown as a
function of momentum (panel (a)) or density (panel (b)) for the MDI
interaction and as a function of density for the SBKD interaction (panel
(c)). Also, evaluating isospin diffusion $R_i$ based on three reaction
systems, that have different time evolutions for the projectile residue as a
result of different total energies and numbers of nucleons, further
contributes to the change of $R_i$ at low density. For the two interactions
consider here, the main difference between the values for $R_i$ appears in
the expansion phase when densities in the participant region are well below $%
\rho _{0}$. The experimental data from MSU are seen to be reproduced nicely
by the MDI interaction with $x=-1$, while the SBKD interaction with $x=-1$
leads to a significantly lower value for $R_{i}$ as the strength of the
momentum-independent potential is stronger (see Fig.\ \ref{Upot}), which has
been shown to enhance the isospin diffusion \cite{farine91,li04prc,tsang04}.

\begin{figure}[th]
\includegraphics[scale=0.75]{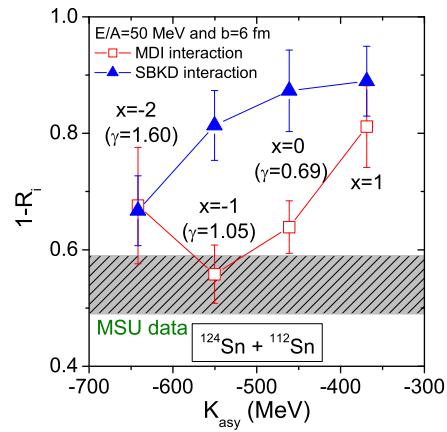}
\caption{{\protect\small (Color online) The degree of isospin diffusion as a
function of $K_{\text{asy}}$ with the MDI and SBKD interactions. $\protect%
\gamma $ is the parameter for fitting the corresponding symmetry energy with 
$E_{\text{sym}}(\protect\rho )=31.6(\protect\rho /\protect\rho _{0})^{%
\protect\gamma }$.}}
\label{RiKasy}
\end{figure}

To see how isospin diffusion depends on the density dependence of nuclear
symmetry energy, we show in Fig.\ \ref{RiKasy} the final saturated value for 
$1-R_{i}$, which measures the degree of isospin diffusion, as a function of $%
K_{\text{asy}}$ for both MDI and SBKD interactions. It is obtained by
averaging the value of $1-R_{i}$ after $120$ fm/c with error bars
corresponding to its dispersion, whose magnitude is similar to the error
band shown in Ref.\cite{tsang04} for the theoretical results from the BUU
model. For the SBKD interaction without momentum dependence, the isospin
diffusion decreases monotonically (i.e., increasing value for $R_{i}$) with
increasing strength of $K_{\mathrm{asy}}$ as the corresponding isovector
potential is mostly positive and decreases with increasing stiffness of $E_{%
\mathrm{sym}}(\rho )$ in the whole range of considered $x$ parameter. The
isospin diffusion is reduced when the momentum-dependent interaction MDI is
used as the momentum dependence weakens the strength of symmetry potential
except for $x=-2$. As seen in Fig.\ \ref{Upot}, the symmetry potential in
the MDI interaction has the smallest strength for $x=-1$ as it is close to
zero at $k\approx 1.5$ fm$^{-1}$ and $\rho /\rho _{0}\approx 0.5$, and
increases again with further hardening of the symmetry energy, e.g., $x=-2$,
when it becomes largely negative at all momenta and densities. The MDI
interaction with $x=-1$ thus gives the smallest degree of isospin diffusion
among the interactions considered in present study and reproduces the MSU
data as already shown in Fig.\ \ref{RiTime}.

The symmetry energy in the MDI interaction with $x=-1$ is $E_{\text{sym}%
}(\rho )=13.0(\rho /\rho _{0})^{2/3}+3.7\rho /\rho _{0}+14.9(\rho /\rho
_{0})^{1.57}\approx 31.6(\rho /\rho _{0})^{1.05}$. 
%Its interaction part can be parameterized as 
%$E_{\text{sym}}^{\mathrm{pot}}(\rho )\approx 18.6(\rho /\rho _{0})^{1.43}$. 
It leads to a value of $K_{\mathrm{asy}}\approx -550$ MeV for the isospin
dependent part of the isobaric incompressibility of asymmetric nuclear
matter, which should be compared to the published constraint of $-566\pm
1350<K_{\mathrm{asy}}<139\pm 1617$ MeV extracted earlier from studying giant
monopole resonances \cite{shlomo93}.

It is worthwhile to mention that if the isoscalar part of the SBKD potential
in Eq. (\ref{Usbkd}) is replaced with the momentum-dependent MDYI potential
of Gale \textit{et al.} \cite{gale90}, which has a similar $K_{0}$ as those
for the MDI and SBKD potentials, the resulting $R_{i}=0.37\pm 0.07$ is much
closer to that obtained with the MDI ($R_{i}=0.44\pm 0.05$) than with the
SBKD ($R_{i}=0.19\pm 0.06$) interaction for $x=-1$. The strongly repulsive
momentum-dependent isoscalar potential thus reduces the effect of isovector
potential on the reaction dynamics. Results from present study are therefore
not much affected by the uncertainty in the momentum dependence of the
isovector potential.

In summary, we have used an isospin- and momentum-dependent transport model
to study isospin diffusion in heavy-ion collisions at intermediate energies.
We find that the diffusion of isospins happens mostly during expansion stage
when the density is below normal nuclear matter density. The momentum
dependence in the nuclear potential plays an important role and affects the
sensitivity of the degree of isospin diffusion to the density dependence of
nuclear symmetry energy. Aside from the uncertainty due to residual effects
of preequilibrium particle emission and cluster formation, the present study
shows that within the context of the mean-field interactions considered here
recent experimental data from NSCL/MSU on isospin diffusion is consistent
with a nuclear symmetry energy that is nearly linear in density at subnormal
densities. This leads to a significantly constrained value of about $-550$
MeV for the isospin-dependent part of the isobaric incompressibility of
isospin asymmetric nuclear matter.

We would like to thank Enzo Greco for helpful discussions. This work was
supported in part by the U.S. National Science Foundation under Grant Nos.
PHY-0098805, PHY-0243571 and PHY0354572 as well as the Welch Foundation
under Grant No. A-1358. LWC was also supported by the National Natural
Science Foundation of China under Grant No. 10105008.


\begin{thebibliography}{99}
\bibitem{oya} {\small K. Oyamatsu \textit{et al.}, 
%I. Tanihata, Y. Sugahara, K. Sumiyoshi, and H. Toki, 
Nucl. Phys. \textbf{A634}, 3 (1998).}

\bibitem{brown} {\small B.A. Brown, Phys. Rev. Lett. \textbf{85}, 5296
(2000).}

\bibitem{hor01} {\small C.J. Horowitz, and J. Piekarewicz, Phys. Rev. Lett 
\textbf{86}, 5647 (2001); Phys Rev. C \textbf{63}, 025501 (2001).}

\bibitem{furn02} {\small R.J. Furnstahl, Nucl. Phys. \textbf{A706}, 85
(2002).}

\bibitem{bethe} {\small H.A. Bethe, Rev. Mod. Phys. \textbf{62}, 801 (1990).}

\bibitem{lat01} {\small J.M. Lattimer and M. Prakash, Astr. Phys. Jour. 
\textbf{550}, 426 (2001); Science \textbf{304}, 536 (2004).}

\bibitem{myers} {\small W.D. Myers and W.J. Swiatecki, Nucl. Phys. \textbf{%
A81}, 1 (1966).}

\bibitem{pomorski} {\small K. Pomorski and J. Dudek, Phys. Rev. C \textbf{67}%
, 044316 (2003).}

\bibitem{ireview98} {\small B.A. Li, C.M. Ko, and W. Bauer, topical review,
Int. Jour. Mod. Phys. E \textbf{7}, 147 (1998).}

\bibitem{ibook} {\small Isospin Physics in Heavy-Ion Collisions at
Intermediate Energies, Eds. Bao-An Li and W. Udo Schr\"{o}der (Nova Science
Publishers, Inc, New York, 2001).}

\bibitem{bom} {\small I. Bombaci, in \cite{ibook}, p.35.}

\bibitem{shlomo93} {\small S. Shlomo and D. H. Youngblood, Phys. Rev. C 
\textbf{47}, 529 (1993).}

\bibitem{youngblood99} {\small D.H. Youngblood \textit{et al.}, 
%H.L. Clark, and Y.W. Lui,
Phys. Rev. Lett. \textbf{82}, 691 (1999).}

\bibitem{pawel02} {\small P. Danielewicz \textit{et al.}, 
%R. Lacey, and W. G. Lynch, 
Science \textbf{298}, 1592 (2002).}

\bibitem{li97} {\small B.A. Li, C.M. Ko, and Z.Z. Ren, Phys. Rev. Lett. 
\textbf{78}, 1644 (1997).}

\bibitem{fra1} {\small B.A. Li and C.M. Ko, Nucl. Phys. \textbf{A618}, 498
(1997).}

\bibitem{fra2} {\small V. Baran \textit{et al.}, 
%M. Colonna, M. Di Toro, and A.B. Larionov,
Nucl. Phys. \textbf{A632}, 287 (1998).}

\bibitem{xu00} {\small H.S. Xu \textit{et al.}, Phys. Rev. Lett. \textbf{85}%
, 716 (2000).}

\bibitem{tan01} {\small W.P. Tan \textit{et al.}, Phys. Rev. C \textbf{64},
051901(R) (2001).}

\bibitem{bar02} {\small V. Baran \textit{et al.}, 
%M. Colonna, M. Di Toro, V. Greco, and M. Zielinska-Pfab\'e, and H.H. Wolter, 
Nucl. Phys. \textbf{A703}, 603 (2002).}

\bibitem{betty} {\small M.B. Tsang \textit{et al.}, Phys. Rev. Lett. \textbf{%
86}, 5023 (2001).}

\bibitem{lis} {\small B.A. Li \textit{et al.}, 
%A.T. Sustich, and B. Zhang, 
Phys. Rev. C \textbf{64}, 054604 (2001).}

\bibitem{li00} {\small B.A. Li, Phys. Rev. Lett. \textbf{85}, 4221 (2000).}

\bibitem{npa01} {\small Radioactive Nuclear Beams, a special volume of Nucl.
Phys. \textbf{A693}, (2001), Ed. I. Tanihata.}

\bibitem{li02} {\small B.A. Li, Phys. Rev. Lett. \textbf{88}, 192701 (2002);
Nucl. Phys. A\textbf{708}, 365 (2002).}

\bibitem{chen} {\small L.W. Chen, V. Greco, C.M. Ko, and B.A. Li, Phys. Rev.
Lett. \textbf{90}, 162701 (2003); Phys. Rev. C \textbf{68}, 014605 (2003);
L.W. Chen, C.M. Ko, and B.A. Li, \textit{ibid} C \textbf{68}, 017601 (2003);
Nucl. Phys. \textbf{A729}, 809 (2003).}

\bibitem{farine91} {\small M. Farine \textit{et al.}, Z. Phys. A\textbf{339}%
, 363 (1991).}

\bibitem{shi03} {\small L. Shi and P. Danielewicz, Phys. Rev. C \textbf{68},
064604 (2003).}

\bibitem{li04prc} {\small B.A. Li, Phys. Rev. C \textbf{69}, 034614 (2004).}

\bibitem{tsang04} {\small M.B. Tsang \textit{et al.}, Phys. Rev. Lett. 
\textbf{92}, 062701 (2004).}

\bibitem{gbd87} {\small C. Gale, G. Bertsch, and S. Das Gupta, Phys. Rev. C 
\textbf{35}, 1666 (1987).}

\bibitem{welke88} {\small G.M. Welke \textit{et al.}, 
%M. Prakash, T. T. S. Kuo, S. Das Gupta, and C. Gale, 
Phys. Rev., C \textbf{38}, 2101 (1988).}

\bibitem{gale90} {\small C. Gale \textit{et al.}, 
%G. M. Welke, M. Prakash, S. J. Lee, and S.
Phys. Rev., C \textbf{41}, 1545 (1990).}

\bibitem{pan93} {\small Q. Pan and P. Danielewicz, Phys. Rev. Lett. \textbf{%
70}, 2062 (1993).}

\bibitem{zhang94} {\small J. Zhang \textit{et al.}, 
%S. Das Gupta, and C. Gale, 
Phys. Rev. C \textbf{50}, 1617 (1994).}

\bibitem{greco99} {\small V. Greco \textit{et al.}, 
%A. Guarnera, M. Colonna, and M. Di Toro,
Phys. Rev. C \textbf{59}, 810 (1999).}

\bibitem{dan00} {\small P. Danielewicz, Nucl. Phys. \textbf{A673}, 375
(2000).}

\bibitem{persram02} {\small D. Persram and C. Gale, Phys. Rev. C \textbf{65}%
, 064611 (2002).}

\bibitem{hod94} {\small P.E. Hodgson, The Nucleon Optical Model, World
Scientific, Singapore, 1994.}

\bibitem{das03} {\small C.B. Das, S. Das Gupta, C. Gale, and B.A. Li, Phys.
Rev. C \textbf{67}, 034611 (2003).}

\bibitem{li04mass} {\small B.A. Li, Phys. Rev. C \textbf{69}, 064602 (2004).}

\bibitem{li04a} {\small B.A. Li, C. B. Das, S. Das Gupta, and C. Gale, Phys.
Rev. C \textbf{69}, 011603 (R) (2004); Nucl. Phys. \textbf{A735}, 563 (2004).%
}

\bibitem{rizzo04} {\small J. Rizzo \textit{et al.}, 
%M. Colonna, M. Di Toro, and V. Greco,
Nucl. Phys. \textbf{A732}, 202 (2004).}

\bibitem{chen04} {\small L.W. Chen, C.M. Ko, and B.A. Li, Phys. Rev. C 
\textbf{69}, 054606 (2004).}

\bibitem{baran04} {\small V. Baran et al., 
%M. Colonna, V. Greco, and M. Di Toro,
Phys. Rep., submitted.}

\bibitem{klug03} {\small J. Klug et al., Phys. Rev. C \textbf{67},
0316001(R) (2003); \textbf{68}, 064605 (2003).}

\bibitem{prakash85} {\small M. Prakash and K. S. Bedell, Phys. Rev. C 
\textbf{32}, 1118 (1985).}

\bibitem{lopez88} {\small M. Lopez-Quelle \textit{et al}., 
%S. Marcos, R. Niembro, A. Bouyssy, and N. V. Giai, 
Nucl. Phys. \textbf{A483}, 479 (1988).}

\bibitem{sbkd} {\small G.F. Bertsch \textit{et al.}, 
%H. Kruse and S. Das Gupta, 
Phys. Rev. C \textbf{29}, 673 (1984).}

\bibitem{rami00} {\small F. Rami \textit{et al.}, Phys. Rev. Lett. \textbf{84%
}, 1120 (2000).}
\end{thebibliography}
\end{document}